\documentclass{article}

\def\pa{\partial}
\def\lapl{\triangle}
\def\grad{\nabla}

\def\xb{{\bf x}}

\def\a{\alpha}
\def\b{\beta}
\def\eps{\varepsilon}
\def\la{\lambda}
\def\de{\delta}
\def\s{\sigma}
\def\F{{\cal F}}
\def\({\left(}
\def\){\right)}
\def\[{\left[}
\def\]{\right]}
\def\<{\langle}
\def\>{\rangle}
\def\.#1{{\dot #1}}
\def\^#1{{\widehat #1}}

\def\~#1{{\widetilde #1}}

\begin{document}






\title{Symmetry of stochastic equations}
\author{Giuseppe GAETA \\
Dipartimento di Matematica, Universit\`a di Milano, \\
v. Saldini 50, I--20133 Milano (Italy) \\
{\it gaeta@mat.unimi.it} }

\maketitle

\section*{Introduction}

Symmetry methods are by now recognized as one of the main tools to attack  {\it  deterministic} differential equations (both ODEs and PDEs); see e.g. \cite{VArn,BlK,Gaeb,Ibr,Olv1,Olv2,Ste,Win}. The situation is quite different for what concerns {\it stochastic} differential equations \cite{LArn,Gar,Gue,Kam,Oks}: here, symmetry considerations are of course quite widely used by theoretical physicists (see e.g. \cite{Bar,HHZ,KPZ} in the context of KPZ theory), but a rigorous and general theory comparable to the one developed for deterministic equation is still lacking, and the attention of the community working on symmetry methods should maybe be called to this.

We would like to quote here early work by Misawa \cite{Mis}, mainly concerned with conservation laws for stochastic systems, and on the work by Arnold and Imkeller \cite{LArn,ArI} for normal forms of stochatic equations.

In the following I will report on some work I have done -- to a large extent in collaboration with N. Rodr\'{\ i}guez Quintero -- on symmetries of stochastic (Ito) equations, and how these compare with the symmetries of the associated diffusion (Fokker-Planck) equations \cite{CicVit,Fin,Ris,StSt}; further details can be found in \cite{SDE1,SDE2}. Part of this was concerned with providing a suitable definition for symmetries of a stochastic equation, so I will also show that the symmetries at the basis of the construction of the KPZ equation are actually recovered within the frame developed here.

\section{Projectable symmetries}

We assume the reader to be fairly familiar with the concept of symmetry of a deterministic differential equation, and the methods to characterize these; so we will just recall some formula of use in the following, mainly to fix notation, identify the class of transformations we are going to consider, and compare with the situation for stochastic equations.

Let us first consider the case of systems of first order ODEs in $R^n$,
$$ {\dot x}^i \ - \ f^i (x,t) \ = \ 0 \ . \eqno(1.1) $$

The general form of Lie-point vector fields is in this case, writing $\pa_i = \pa / \pa x^i$,
$ X = \tau (x,t) \pa_t + \xi^i (x,t) \pa_i$.
Such a vector field, however, is somehow too general: indeed, we would like the change on $t$ not to depend on the value assumed by the solution $x(t)$; this is particularly relevant when we think of the ODE as describing not the motion of a single particle in $R^n$, but that of an assembly of particles. We would thus like to keep the special role of time, i.e. require $\tau = \tau (t)$; in other words, we will consider
$$ X_0 \ = \ \tau (x,t) \pa_t + \xi^i (x,t) \pa_i \ . \eqno(1.2) $$

A similar discussion holds for symmetries of PDEs, first or higher order. Here we will deal only with scalar PDEs, i.e. equations for a single dependent variable $u = u(x,t)$. In this case we would write in general
$ X = \tau (x,t,u) \pa_t +  \xi^{i} (x,t,u) \pa_i +  \phi (x,t,u) \pa_u$.
On physical terms, it is preferable to consider only transformations such that the independent variables are transformed independently of the values assumed by the dependent ones, i.e. such that $\tau$ and the $\xi^i$ do not depend on $u$; also, here we are primarily interested in evolution equations, and in order to guarantee the time keeps its distinguished role we should also ask that $\tau$ does not depend on $x$. Thus, in the end we want to consider transformations of the
form
$$ X \ = \ \tau (t) {\pa \over \pa t} +  \xi^{i} (x,t) {\pa \over \pa x^{i} } +  \phi (x,t,u) {\pa \over \pa u } \ . \eqno(1.3) $$

\medskip\noindent
{\bf Definition.} {\it The vector fields -- and in particular symmetry generators -- of the form (1.2), (1.3) will be called ``fiber-preserving'', or simply {\bf projectable}. }
\medskip

Let us now consider in particular a system of ODEs, which we write in the form (1.1).
The determining equations for the symmetry generators of this are easily determined (these should be compared to the determining equations for stochastic differential equations discussed in the next section).
We will use, here and below, the notation  $$ \{ f , \xi \}^i \ := \ (f^j \cdot \pa_j)
\xi^i - (\xi^j \cdot \pa_j)  f^i \ . \eqno(1.4) $$

\medskip\noindent
{\bf Lemma 1.} {\it The projectable symmetries of (1.9) are given by vector fields $X_0$ as above with coefficients  satisfying
$$ \pa_t (\xi^i - \tau f^i ) + \{ f , \xi \}^i \ = \ 0 \ . \eqno(1.5) $$}
\medskip

For given functions $f^{i}(x,t)$ the partial differential equations (1.16) always \cite{Olv1,Ste} have non-trivial solutions $\tau(t)$ and $\xi^{i}(x,t)$,  for example we can fix $\tau(t)$ or $\xi^{i}(x,t)$ and compute $\xi^{i}(x,t)$ or $\tau(t)$. Although an infinite number of symmetries of (1.16) exist, in general there is no
constructive way to find them.

For further detail on symmetries of (deterministic) ODEs and PDEs, and
applications, the  reader is referred to \cite{VArn,BlK,Gaeb,Ibr,Olv1,Olv2,Ste,Win}.

\section{Stochastic differential equations}

Let us turn to stochastic equations \cite{LArn,DWM,Gar,Gue,Kam,Oks}, and let us consider an Ito
equation\footnote{It is well known that any Ito equation  is equivalent
to a Stratonovich equation; for the Ito equation (2.1), the associated
Stratonovich equation is
$ dx^i = b^i (x,t) dt + \s^i_k (x,t) \circ dw^k$, with
$ b^i = f^i - (1/2) [ \s (\pa \s^T / \pa x)]$.
We will to the Ito formulation in all of this note.}
$$ dx^i \ = \ f^i (x,t) \, dt \ + \ \s^i_k (x,t) \, d w^k \ , \eqno(2.1) $$
where $f$ and $\s$ are smooth functions,
$\sigma(x,t)$ is a nonzero matrix and the $w^k$ are independent
homogeneous standard Wiener processes, so that
$$ \< | w^i (t) - w^j (s) |^2 \> = \delta^{ij} (t-s) \ .  $$
The equation (2.1) should be seen as a map from the vector Wiener process
${\bf{w}}(t) $ $ = $ $ \{ w^1 (t) , $ $... ,$ $w^n (t) \}$ to the
stochastic process undergone by
$\{ x^1 (t) , ... , x^n (t) \}$, and its meaning is precisely that of
defining the vector stochastic process ${\bf x} (t)$.

If we consider a function $y = \Phi (x)$, its evolution under the Ito
equation (2.1) is described (again in terms of a stochastic equation) by the {\it Ito formula}:
$$ \begin{array}{rl}
dy^i = & {\pa \Phi^i \over \pa x^j }\, d x^j + {1 \over 2}
{\pa^2 \Phi^i \over \pa x^j \pa x^k} \, dx^j dx^k \\
& = \[ f^{j} \pa_j \Phi^i +
(1/2) \pa^2_{jm} \Phi^i (\s \s^T)^{jm} \] \, dt \ + \ \[ (\pa_j \phi^i )
\s^{jk} \] dw^k \ . \end{array} \eqno(2.2) $$

To the Ito equation (2.1) is associated the corresponding Fokker-Planck
equation
$$ \pa_t u \ = \ - \, \pa_i \, (f^i u ) \ + \ {1 \over 2} \,
\pa_{ij}^2 \, \[ (\s \s^T )^{ij} \, u \] \ , \eqno(2.3) $$
describing the evolution of the probability measure $\rho = u (x,t)$ under
the stochastic process described by (2.1). Equations
(2.1) and (2.3) contains the same {\it statistical}
information \cite{LArn,Gar,Gue,Kam,Oks} as far as one-particle processes are concerned
(but not if we think the Ito equation as describing a random dynamical system, see \cite{LArn,SDE2}),
provided $\s$ satisfies the {\bf non-degeneracy condition }
$$ A \ := \ {1 \over 2} \ \s \, \s^T \ \not= \ 0 \, $$
which we will assume throughout this note.

We stress that we are interested in equation (2.3) as far as it describes the time
evolution of the probability measure $\rho = u(x,t)$ under the stochastic
process (2.1). It is obvious that for this interpretation $u(x,t)$
should be non-negative and subject to the condition
$$ \int_{-\infty}^{+\infty} \, u (x,t) \, dx^1...dx^n \ = \ 1 \ . \eqno(2.4) $$

This is relevant in connection with
the allowed transformations in the $(x,t;u)$ space: only
transformations preserving this normalization do represent symmetries of
the Fokker-Planck equation compatible with its probabilistic interpretation\footnote{In
other words, we should set the FP equation in the function space $\F$ defined by (2.4), and thus accept only transformations mapping $\F$ to itself.}, and
one should expect a correspondence between symmetries of the Ito
equation and this subclass -- rather than all -- of the symmetries of the
Fokker-Planck equation.

Another simple but relevant point which should be stressed is the following:
different Ito equations which have the same $f$ but different matrices $\s$ can provide the same term $\s \s^T$ and thus the same Fokker-Planck equation.

A simple example is provided e.g. by $\s$
orthogonal: in this case we have by definition $\s \s^T
= I$, so that all the Ito equations with the same $f$ and any
orthogonal matrix $\s$ give the same Fokker-Planck
equation\footnote{The one-point stochastic processes described by two different Ito equations having
the same associated Fokker-Planck equation have the same statistical
properties (the probability measures evolve in the same way), but {\it
are } different: the same realization of the Wiener process ${\bf w}(t)$
leads to different sample paths.}.
Similarly, $\s$ and $\~\s = \s B$, with $B$ any orthogonal matrix,
will give the same Fokker-Planck equation (and conversely $\s$ and $\~\s$
give the same Fokker-Planck -- with the same $f$ -- only if there is an
orthogonal matrix $B$ such that the above relation is satisfied.

Thus if we consider a continuous variation of $\s$, say $\s + \eps \gamma$,
in the Ito equation, the associated Fokker-Planck equation remains
unchanged provided $(\s + \eps \gamma ) (\s + \eps \gamma )^T = \s
\s^T$, which at order $\eps$ is simply
$$ \s  \gamma^T + \gamma \s^T \ = \ 0 \ . \eqno(2.5) $$

\section{Symmetries of stochastic ODEs}

\subsection{Spatial symmetries}

We will at first consider symmetries not acting on the time variable $t$, i.e. $X_0 = \xi^i (x,t) \pa_i$.
Thus, we  consider a near-identity change of coordinates, passing from
$x$ to $y$ via
$$ x^i \to y^i = x^i + \eps \xi^i (x,t) \ . \eqno(3.1) $$
Using the Ito formula, we easily check that at first order in $\eps$,
this transformation maps the Ito equation (2.1) into a new Ito equation
$$ d y^i \ = \ \~f^i (y,t) \, dt \ + \ \~\s^i_k (y,t) dw^k \ ,  \eqno(3.2) $$
where we have explicitly
$$ \begin{array}{rl}
\~f^i = \ & f^i \ + \ \eps \, [ \pa_t \xi^i + f^j \pa_j \xi^i - \xi^j
\pa_j f^i + (1/2) (\s \s^T)^{jk} \pa^2_{jk} \xi^i ]  \ \\
\~\s^i_k = \ & \s^i_k \ + \ \eps \, [ \s^j_k \pa_j \xi^i  - \xi^j
\pa_j \s^i_k ] \ . \end{array} \eqno(3.3) $$

When (3.1) maps (2.1) into itself, i.e. when (3.2)
coincides with (2.1) [up to terms $o (\eps )$], we say that $X_0 = \xi^i (x,t) \pa_i$ is a
(Lie-point) spatial {\bf symmetry} of (2.1). Thus, Lie-point spatial symmetries
of (2.1) are identified by the vanishing of terms $O (\eps )$ in (3.3). In the following statement, we use the notations introduced above for $A$ and $\{.,.\}$.

\medskip\noindent
{\bf Proposition 1.} {\it The (generators of) Lie-point spatial symmetries of the Ito equation (2.1) are given by vector fields $X_0 = \xi^i (x,t) \pa_x^i$ with coefficients $\xi^i$ satisfying the {\rm determining
equations for spatial symmetries} of an Ito equation:
$$ \begin{array}{l}
\pa_t \xi^i + \{ f,\xi \}^i + (1/2) (\s \s^T )^{jk} \pa^2_{jk} \xi^i = 0 \\
\{ \s_k , \xi \}^i = 0 \ . \end{array} \eqno(3.4) $$}
\medskip

These are $n+n^2$ equations for the $n+1$ functions $\xi^i , \tau$;
thus for $n>1$ they have some solution only in very exceptional cases.
This should not be surprising, as symmetry is a non-generic property.

Note that if $\s = \s (t)$ does not depend on the spatial
variables, the second of (3.4) reduces to $\s^j_k \pa_j \xi^i = 0$,
which in turns imply the vanishing of the term $(\s \s^T )^{jk}
\pa^2_{jk} \xi^i$, which in this case can be rewritten as $\s^k_p \pa_k
(\s^j_p \pa_j \xi^i)$. Thus, (3.4) are in this case equivalent to the
determining equations for symmetries of the deterministic part of the
Ito equation [see (1.5) and recall now we are assuming $\tau = 0$]
with the additional condition $\s^j_k \pa_j \xi^i = 0$.

\subsection{Spatio-temporal symmetries.}

We could of course consider transformations acting also on $t$: in this
case the $t$ transformation would also affect the processes $w (t)$,
and some extra care should be paid. We will not go over the details (which consist
in studying how $w (t)$ is changed and include this in the change undergone by $\s$), and just report the result obtained in \cite{SDE1}, with the same notation as above.

\medskip\noindent
{\bf Proposition 2.} {\it The projectable vector field $X_0 = \tau (t)
\pa_t + \xi^i (x,t) \pa_x^i$ is a symmetry generator for
the Ito equation (2.1) if and only if the coefficients $\{\tau, \xi^i\} $
satisfy the {\rm full determining equations} for projectable symmetries of an Ito equation:
$$ \begin{array}{l}
\pa_t \( \xi^i - \tau f^i \)  +  \{f,\xi\}^i - A^{jk}
\pa^2_{jk} \xi^i \ = \ 0 \\
\{ \s_k , \xi \}^i - \tau \pa_t \s^i_k - (1/2) \s^i_k \pa_t \tau \
= \ 0 \ . \end{array} \eqno(3.5) $$}

Note that the symmetries which are linear in $x$, i.e. such that
$\xi^i (t;{\bf x}) = M^i_j (t) x^j$, are given by the same equations as
for symmetries of the deterministic part of the Ito equation (2.1),
i.e. (1.5), plus the additional condition
$ M^i_j \s^j_k = M^j_p x^p \pa_j \s^i_k + \tau \pa_t \s^i_k + (1/2)
\s^i_k \pa_t \tau$.

A relevant case in applications is the one where $f^i =
f^i(\xb)$ and $\s^i_k = \s^i_k (t)$ or even $\s^i_k = {\rm const} =
S^i_k$; this correspond to an autonomous dynamical system subject to a
noise which depends only on $t$ or even a constant noise. In this case
(3.4) can be discussed quite completely. Indeed, with $\s$
independent of the spatial coordinates, the second of these reads
$ 2 \s^j_k \pa_j \xi^i = 2 \tau \pa_t \s^i_k + \s^i_k \pa_t \tau $.
As the r.h.s. only depends on $t$, by differentiating with respect to
$x^m$ we get the equation $ \s^j_k (\pa^2 \xi^i / \pa x^j \pa x^m)  =  0$:
i.e., the (symmetric) matrix of second derivatives of $\xi^i$, $H^i_{jm} $
must be such that $\s H^i = H^i \s = 0$.
Notice that in particular, if $\s^i_k (t) = \la_{(i)} (t) \delta^i_k $
with all $\la_{(i)} \not\equiv 0$ (or however if $\s^{-1}$ exists),
this means that $\xi$ can be at most linear in the $x$.

\section{Symmetries of the Fokker-Planck equation.}

We will now discuss the determining equations for projectable
symmetry generators of the Fokker-Planck equation (2.3) in arbitrary
spatial dimensions; general
symmetries of the Fokker-Planck equations (with some limitations on
$\s$) in one and two space dimensions have been completely classified
\cite{CicVit,Fin,StSt}.

It will be convenient to rewrite the Fokker-Planck equation as
$$ u_t \ + \ A^{ij} \, \pa^{2}_{ij} u \ + \ B^i \, \pa_{i} u  \ + \ C
u \ = \ 0 \ ;
\eqno(4.1) $$
the coefficients $A,B,C$ depend on $x$ and $t$ only
and are given explicitly by\footnote{Recall that our non-degeneracy assumption guarantees $A \not= 0$.}
$$ A =  -(1/2) \s \s^T  \ ; \
B^i = f^i + 2 \pa_j A^{ij} \ ; \
C (x,t) = (\pa_i \cdot f^i ) + \pa^2_{ij} A^{ij} \ . \eqno(4.2) $$

We consider a projectable vector field of the form (1.3).
Computing the second prolongation of this and applying it to the equation,
we obtain the determining equations for symmetries of the Fokker-Planck equation.

It follows by general results on Lie-point symmetries of linear equations \cite{BlK,Gaeb,Olv1}, or by an elementary explicit computation, that:

\medskip\noindent
{\bf Lemma 2.} {\it The projectable symmetries of the Fokker-Planck
equation (4.1) are given by vector fields in the form (1.8) with
$\phi = \a (x,t) + \b (x,t) u$.}
\medskip

Again by linearity, we will have ``trivial'' symmetries $X_\a = \a (x,t) \pa_u$,
with $\a (x,t)$ an arbitrary solution to the FP equation itself; this is just
expressing the linear superposition principle \cite{Olv1}. Note that in order to preserve the normalization (2.4), we should moreover require $\int \a (x,t) d x = 0$, which rules out nontrivial non-negative solutions.

We will thus from now on focus on other, i.e. ``nontrivial'', symmetries only, and take $\a (x,t) = 0$.

\medskip\noindent
{\bf Proposition 3.} {\it The nontrivial projectable symmetries of the Fokker-Planck equation (4.1) are given by vector fields in the form (1.3) with
$\phi = \b (x,t) u$, where $\tau$, $\xi^i$, $\b$ satisfy the {\rm determining equations}
$$ \begin{array}{l}
\pa_t (\tau A^{ik} ) + (\xi^m \pa_{m} A^{ik} - A^{im} \pa_{m}
\xi^k - A^{mk} \pa_{m} \xi^i ) = 0 \\
\pa_t (\tau B^i ) - \[ \xi^i_t + B^m \pa_{m} \xi^i - \xi^m \pa_{m} B^i  \] + ( A^{ik} \pa_{k} \b + A^{mi} \pa_{m} \b ) - A^{mk} \pa_{mk}^{2} \xi^i = 0 \\
\pa_t (\tau C ) + \b_t + A^{ik} \pa_{ik}^{2} \b + B^i \pa_{i} \b +  \xi^m
\pa_{m} C = 0 \ . \end{array} \eqno(4.3) $$}
\medskip

It should be stressed that in the computations leading to (4.3) we have not
required preservation of the normalization $\int u \ d x = 1$; when
we require this, it turns out that

\medskip
{\bf Lemma 3.} {\it The nontrivial projectable symmetries of the Fokker-Planck
equation (4.1), in the form (1.3) with $\phi = \b (x,t) u$,
preserve the normalization (2.4) if and only if
$$ \b \ = \ - {\rm div} (\xi) \ . $$}
\medskip

In other words, this lemma guarantees that if we are looking for the (nontrivial) symmetries of a Fokker-Planck equation compatible with its probabilistic interpretation, we do not have to deal with the general form (1.3)
of the symmetry vector field, but we can use instead the ansatz
$$ X \ = \ \tau (t) \pa_t \, + \, \xi^i (x,t)
\pa_i \, - \, [{\rm div} (\xi) u] \pa_u \ ; \eqno (4.4) $$
this substantially simplifies the analysis of symmetries.

\section{Symmetries of the Ito versus
symmetries of the associated FP equation.}

We are specially interested in discussing how the symmetries of the
partial differential equation (4.1) and those of the symmetries of the
system of stochastic ODEs (2.1) are related.

In order to do this, we should express the coefficients $A,B,C$ in terms of
$f$ and $\s$; indeed $A = - (1/2) \s \s^T$, $ B^i = f^i + 2 \pa_k A^{ik}$,
and $C = \pa_i f^i + \pa^2_{ik} A^{ik}$.
Using the latter two of these and with some manipulations, see \cite{SDE1},
(4.3) reads
$$ \begin{array}{l}
\pa_t (\tau A^{ik} ) + \( \xi^m \pa_{m} A^{ik} - A^{im} \pa_{m} \xi^k -
A^{km} \pa_{m} \xi^i \) \ = \ 0 \\
\[ \pa_t ( \xi^i - \tau f^i )  + \{ f, \xi \}^i - A^{mk} \pa^{2}_{mk}
\xi^i \]  -  2 \[ A^{ik} \pa_{k} \b + \ A^{im} \pa_{mk}^{2} \xi^k
 \] \ = \ 0 \\
\[ \pa_t  + f^i \pa_{i} - A^{ik} \pa^{2}_{ik} \] \[ \b +  \pa_{m} \xi^m
\] \ = \ 0 \ . \end{array} \eqno(5.1) $$

It is convenient to introduce the shorthand notations
$$ \begin{array}{l}
\Gamma^k_j \ = \ \s^m_j \pa_m \xi^k - \xi^m \pa_m \s^k_j -
\tau \pa_t \s^k_j - (1/2) \s^k_j \pa_t \tau \ , \\
\Lambda^i \ = \ - \, \[ \pa_t ( \xi^i - \tau f^i )  + \{ f, \xi \}^i - A^{mk} \pa^{2}_{mk}
\xi^i \] \ . \end{array} \eqno(5.2) $$
With these, (5.1) is rewritten as
$$ \begin{array}{l}
\s^i_j \Gamma^k_s \delta^{js} + \s^k_j \Gamma^i_s \delta^{js} = 0 \\
\Lambda^i + 2 \[ A^{ik} \pa_{k} \b + \ A^{im} \pa_{mk}^{2} \xi^k \] \\
\[ \pa_t  + f^i \pa_{i} - A^{ik} \pa^{2}_{ik} \] \[ \b +  \pa_{m} \xi^m
\] \ = \ 0 \ . \end{array} \eqno(5.3) $$

It should be stressed that with the notation just introduced, the determining equations (3.5) for symmetries of the Ito equation (2.1) read simply
$$ \Lambda^i \ = \ 0  \ \ ; \ \ \Gamma^i_j \ = \ 0 \ .  \eqno(5.4) $$

With this notation, the comparison of symmetries for the Ito and the associated Fokker-Planck equation is quite a simple matter.

\subsection{From the Ito to the FP symmetries}

We will first investigate if symmetries of an Ito equation result in symmetries
of the associated Fokker-Planck equation.

We first note that for $\Gamma = 0$, the first of (5.3) is satisfied.
As for the second of (5.3), when $\Lambda^i = 0$ this reduces to
$$ A^{ik}\pa_k \b = - A^{im} \pa^2_{mk} \xi^k \ . \eqno(5.5) $$

Obviously $\b$ is not present in symmetries of the Ito equation, so we
can choose it as to satisfy the third of (5.3) and (5.5): for this it
suffices to choose
$$ \b \ = \ - \pa_m \xi^m + c_0 \ = \ - {\rm div} (\xi) + c_0 \ .
\eqno(5.6) $$
We know from lemma 2 that actually we have to take $c_0 = 0$, and that this is actually the only choice of $\b$ providing us with a normalization-preserving symmetry of the FP. Our findings are summarized as follows:

\medskip
{\bf Proposition 4.} {\it Let $X_0 = \tau \pa_t + \xi^i \pa_i$ be a symmetry of the Ito equation (2.1). Then $X_0$ extends to a unique normalization preserving symmetry $X := X_0 - [({\rm div} \xi) u] \pa_u$ of the associated Fokker-Planck equation.}
\medskip

\subsection{From the FP to the Ito symmetries}

Let us now consider the converse question, i.e. if and when a
symmetry of the FP equation associated to an Ito equation can
be projected to a symmetry of the Ito equation itself.

As the probabilistic interpretation of the FP requires the normalization
condition (2.4), we expect only symmetries identified by lemma 2 can provide
symmetries of related Ito equations.

Moreover, as recalled above, there are transformations which map
an Ito equation into a different one with the same statistical properties
and thus the same associated FP equation: these would be
symmetries of the FP but not of the Ito equation. Thus, not all
the symmetries of the FP should be expected to produce symmetries of the
Ito equation.

By lemma 2, we assume $\b = - {\rm div} (\xi)$; this guarantees
that the third of (5.3) holds, and also that the second of (5.3)
reduces to $\Lambda^i = 0$.
Hence we only have to discuss the relation between the first of (5.3) and the
second of (5.4).
These two equations can be rewritten, respectively, as
$$ \s \Gamma^T + \Gamma \s^T = 0 \ \ {\rm  and} \ \ \Gamma = 0 \ . \eqno(5.7) $$
The second of these imply the first, but the converse is not true:
hence there is {\it not} a complete equivalence.

This just corresponds to a given FP equation corresponding to different Ito equations: the transformations with $\Gamma \not= 0$ but $\s \Gamma^T + \Gamma \s^T = 0$ will be precisely those which map an Ito equation $E_0$ into a different Ito equation $E_1$ which has the {\it same} Fokker-Planck associated
equation, see (2.5).

\medskip
{\bf Proposition 5.} {\it Let $X = \tau (t) \pa_t +
\xi^i (x,t) \pa_i - [({\rm div} \xi) u] \pa_u$ be a
symmetry of the Fokker-Planck equation associated to the Ito equation (2.1).
Then $X$ maps the Ito equation into a (generally, different) Ito equation
with the same statistical properties; $X$ is a symmetry of the Ito
equation (2.1) if and only if $\Gamma$ defined in
(5.4) satisfies $\Gamma^i_k = 0$.}
\medskip

We may note that from the above system (5.4), restricting to the case
where  $f(t,x)=f(x)$ and $\s(t,x)=\s(x)$, we recover the results of
Cicogna and Vitali \cite{CicVit} for the one-dimensional setting. From (5.2)
it is also possible to recover the results of Shtelen and Stogny
\cite{StSt} for the two-dimensional Kramers equation, as well as the
results of Finkel \cite{Fin} for certain two-dimensional FP equations.

\subsection{Examples}

{\bf Example 1.} As the first example, we consider
the case $f(t,x) = 0$, $\s(t,x) = \s_0 = {\rm const} \not=0$, i.e. the
equation $$ dx = \s_0 \, dw(t) \eqno(5.8) $$
which represents a free particle subject to constant noise. The
corresponding Fokker-Planck equation is simply the heat equation
$ u_t = (\s_0^2 / 2) u_{xx}$.
The symmetries of the heat equation (other than the trivial ones
$v_\a = \a (x,t) \pa_u$ with $\a_t = \a_{xx}$)
are well known to be \cite{Olv1,Ste}
$$ \begin{array}{l}
v_1 =  \pa_t \ ; \
v_2 =  \pa_x \ ; \
v_3 =  u \pa_u \ ; \
v_4 =  \s_0^2 t \pa_x - \s_0 x u \pa_u \ ; \\
v_5 =  2 t \pa_t + x \pa_x \ ; \
v_6 =  t^2 \pa_t + x t \pa_x - (1/2) (t + x^2 / \s_0^2 ) u \pa_u \end{array} \eqno(6.2) $$
Of these, $v_1 , v_2$ and $v_5$ (which do not act on $u$) are also
symmetries of the Ito equation (5.8), as is easily checked using (3.4).
Notice that (4.4) is satisfied for these, and is not satisfied for
$v_3,v_4$ and $v_6$. The vector fields $v_1 , v_2$ and $v_5$ do actually span the symmetry algebra of (5.8).
\bigskip

{\bf Example 2.} As an example in two space dimensions [with
coordinates $(x_1,x_2) = (x,y)$], we choose
$$ \begin{array}{rl}
dx \ = \ & y \, dt \\
dy \ = \ & - k^2 y  \, dt \ + \
\sqrt{2 k^2} \, dw(t) \  \end{array} \eqno(5.9) $$
with $k^2$ a positive constant.

The corresponding Fokker-Planck equation is the Kramers equation
$$ u_t \ = \ k^2  u_{yy} - y u_x  + k^2 y u_y +
k^2 u \ ; \eqno(5.10) $$
the symmetries of this were studied in \cite{StSt} and, apart from
the trivial ones $v_\a$, are
$$ \begin{array}{l}
v_1 \ =  \pa_t \ ; \
v_2 \ =  \pa_x \ ; \
v_3 \ =  e^{-k^2 t} \, [ k^{-2} \pa_x - \pa_y ] \ ; \
v_4 \ =  u \pa_u \ ; \\
v_5 \ =  t \pa_x + \pa_y - (1/2) (y + k^2 x) u \pa_u \ ; \
v_6 \ =  e^{k^2 t} \, [ k^{-2} \pa_x + \pa_y - yu \pa_u ] \end{array} $$
Here $v_1 , v_2$ and $v_3$ satisfy (4.4), while for $v_4 , v_5$ and
$v_6$ this is violated.
According to our definition, the symmetries of the equations
(5.9) are again $v_1$, $v_2$ and $v_3$, as easily checked.
\bigskip

{\bf Example 3.} We consider now an example where the correspondence between
normalization-preserving symmetries of the FP equation and
symmetries of the Ito equation is not complete, i.e. the two-dimensional Ito system (with zero drift)
$$ \begin{array}{rl}
dx^1 \ = & \ \cos (t) dw^1 \, - \, \sin (t) d w^2   \\
dx^2 \ = & \ \sin (t) dw^1 \, + \, \cos (t) dw^2 \ ; \end{array}  \eqno(5.11) $$
the corresponding Fokker-Planck equation is now just the two dimensional heat
equation $u_t = (1/2) \triangle u$.

It is now immediate to check that the vector field $X_0 = \pa_t$ is a
symmetry of the FP equation, but {\it not } a symmetry of (5.11).
Obviously, the case of any orthogonal $\s$ with $\pa_t \s^i_k \not\equiv 0$
will be exactly the same.
\bigskip

{\bf Example 4.} Consider $n$ uncoupled equations
for equal ``Langevin harmonic oscillators''
subject to independent stochastic noises
\cite{Kam,Ris}; this system is described by the Ito system
$$ dx^{i} \ = \ - x^{i} \, dt \, + \, \sqrt{2 s_i} \, dw^i \ ,
\ \ i=1, 2, ..., n \, \eqno(5.12) $$
(no sum on $i$) where we assume all the $s_i$ are strictly
positive; the corresponding Fokker-Planck equation is
$$ \pa_{t} u \ = \ \sum_{i=1}^n \, [ s_i \pa^{2}_{ii} u +  x^{i}  \pa_{i} u + u ] \ .  $$
In this case we get the symmetries ($i=1,2,...,n$)
$$ v_1 = \pa_t \ , \
v_{2} = e^{-2 t}  [ \pa_t - \sum_{i=1}^{n} x^{i} \pa_i + n u \pa_u ] \ , \
v_{q_i} = e^{-t} \pa_i \ .  $$
\bigskip

{\bf Example 5.} Finally, we consider now the n-dimensional nonlinear
case
$$ dx^{i} = -  (1 - \lambda \| x \|^2) x^{i} \, dt + dw^{i} \ , \eqno(5.13) $$
where $ \| x \|$ is the norm of the vector ${\bf x}$ and $\lambda \not= 0$.
By inserting $f^{i}= - x^{i}(1 - \lambda \| x \|^2)$ and
$\s_{j}^{i} = \delta_{ij}$ in (5.4) we obtain, from the second of these,
$\xi^i = h^i (t) + (\tau_t/2) x^i$.
Inserting this into the first of (5.4) and isolating the coefficients
of different powers of $x$, we get that the only symmetry is given by $v = \pa_t$.

Note that we got no rotation symmetry; this is because within the class of
transformations we are considering we can rotate the vector $x$, but {\it not} the vector Wiener process $w(t)$; transformations allowing to rotate $w(t)$ as
well will be considered in the next section.

\section{W-symmetries}

We will consider symmetries involving not only the spatial and time
variables $(x , t)$, but also the vector Wiener processes $w (t)$
entering in the $n$-dimensional Ito equation (2.1).
We will specifically consider infinitesimal transformations of the form
$$ \begin{array}{l}
x^i \to  y^i \ =  x^i + \eps \xi^i (x,t) \\
t  \to  s \ =  t + \eps \tau (t) \\
w^i \to  z^i \ =  w^i + \eps \mu^i (w,t) \ . \end{array} \eqno(6.1) $$
We also call symmetry generators of this form, ``{\it W-symmetries}''.

Note that this is not the most general possible form of a transformation
for the variables involved in (1.1); some words on this restriction are in order here.

The restriction on $\tau$ and $\xi$ is the same as considered above.
Moreover, we only allow ``internal'' transformations of $w$, i.e. they cannot depend on $x(t)$.  As we think of the stochastic process $w (t)$ to be
independent of the evolution of the $x (t)$, we like its transformation not to depend on the latter.
Finally, we have allowed the transformation on the spatial coordinates
$x(t)$ to depend on $x$ and $t$, but not on the $w$; this means
that we do not want to consider transformations of the spatial
coordinates which depend on the realization of the stochastic process
$w (t)$. This again is somewhat a natural requirement in physical terms\footnote{It should be however stressed that in this way we are also discarding the transformations needed to obtain {\it normal forms} of stochastic
differential equations, see \cite{LArn,ArI} and references therein.}.

In order to compute how (6.1) acts on the Ito equation (2.1), we use the Ito formula (2.2). Moreover, due to the form of (6.1), we can adopt a ``two-steps procedure'', see \cite{SDE2}: the transformations of the $(x,t)$ variables and of the stochastic process $w(t)$ do not interfere with each other. It turns out that some strong limitation on the functions $\mu^i$ arise.

\medskip\noindent
{\bf Lemma 4.} {\it The infinitesimal transformation $w \to z = w + \eps \mu (w,t)$ maps the vector Wiener process $w(t)$ into a vector Wiener process $z(t)$ if and only if $\mu = B w$, with $B$ a real antisymmetric matrix. Equivalently, if and only if $z = M w$ with $M$ orthogonal.}
\medskip

Note that as $B$ is antisymmetric matrix, it vanishes in dimension one:
for one-dimensional Ito equations we have no new symmetries with respect
to those discussed above.

\medskip\noindent
{\bf Proposition 6.} {\it Under the infinitesimal transformation (6.1), with $B$ a real antisymmetric matrix, the Ito equation (2.1)
is changed into a (generally, different) Ito equation. The Ito equation (2.1) is invariant under (6.1) if and only if $\tau,\xi,\mu$ satisfy the {\it determining equations for W-symmetries}:
$$ \begin{array}{l}
\pa_t \xi^i \, + \, \{ f , \xi \}^i  \, - \,
\pa_t (\tau f^i) \, + \, A^{jk} \pa^2_{jk} \xi^i \ = \ 0 \\
\{ \s_k , \xi \}^i - \tau \pa_t \s^i_k - (1/2)
(\pa_t \tau) \s^i_k - \s^i_p B^p_k \ = \ 0 \ . \end{array} \eqno(6.2) $$}
\medskip

If we just consider a transformation $w \to z = M w$, with $M$ orthogonal as required by lemma 4, the Fokker-Planck equations associated to the Ito equation and to the transformed one will be the same. Thus all the discussion conducted in
section 5 about the relations betwen symmetries of an Ito
equation and of the associated FP equation also applies to transformations of the form (6.1).\footnote{Needless to say, for more general transformations, for which the transformations of the $(x,t)$ variables and of the vector Wiener process
$w(t)$ do interfere with each other, this would not be the case.}

\bigskip

{\bf Example 1.} Let us consider again the two-dimensional Ito equation (5.9)
If we look at the second set of determining equations and single out the one for $i=2$ and $k=1$, we have immediately that $B=0$, i.e. in this case we have no new symmetry by allowing transformation of $w$.
\medskip

{\bf Example 2.} Let us consider again the two-dimensional Ito system (5.11)
We will write $\xi^1 = \a$, $\xi^2 = \b$, and $B_{12}=-B_{21} = b$.
The most general solution to the determining equations turns out to be
$$ \begin{array}{l}
\a \ = \ [\tau_t / 2 - \sin (2t) \tau ] x + [ b + \cos (2t) \tau ] y +
\a_0 (t) \\
\b \ = \ - [b + \tau ] x + [\tau_t / 2] y + \b_0 (t) \ ; \end{array} $$
By setting $\tau = \a_0 = \b_0 = 0$, we obtain a new symmetry,
corresponding to $\a = b y$ and $\b = - b x$ with
$b$ arbitrary, i.e. given by
$$ X \ = \ \( y {\pa \over \pa x} \, - \, x {\pa \over \pa y} \)  \ + \
\( w^2 {\pa \over \pa w^1} \, - \, w^1 {\pa \over \pa w^2} \) \ . $$
Note this is nothing else than a simultaneous (and identical)
rotation in the $(x,y)$ and in the $(w_1 , w_2 )$ planes.
\medskip

{\tt Example 3.} We will now consider again the case of $n$ uncoupled equal ``Langevin harmonic oscillators'', see (5.12) above.

If we discard symmetries with $B=0$ (i.e. those obtained above), we are
still left with a nontrivial possibility, i.e. $B$ an arbitrary (real,
antisymmetric) constant matrix and $\xi^i = C^i_{~k} x^k$, where
$C^i_{~k} = (s_i / s_k ) B^i_k$ (no sum on $i,k$ here).

The meaning of this result is obvious: we can act on this system by an
arbitrary $SO(n)$ rotation in the $w$ space and a
related rotation in the $x$ space.

If we assume $s_1 = s_2 = ... = s_n = s$, we have indeed $C=B$ and the
rotations in $w$ and $x$ spaces do just coincide. For general $s_1
, ... , s_n$, the $(s_i/s_k)$ factors relating the $C$ and $B$ matrices
are also easily understood: we could rescale each of the $x^i$ by a
factor $\sqrt{2 s_i}$, $x^i = \sqrt{2 s_i} y^i$ (no sum on $i$),
arriving at the manifestly rotationally invariant $n$-dimensional Ito
equation $d y^i = - y^i d t + d w^i$.
\medskip

{\bf Example 4.} Let us consider again (5.13).
We easily check that it is rotationally invariant with a W-symmetry:
take $\tau = 0$ and $\xi^i = B^i_k x^k$, so that
$$ \{ \xi , f \}^i \ = \ \{ B x , (1 - \lambda ||x||^2 ) x \} \ =
\  - 2 \lambda x^i \( x^j B^j_k x^k \) \ = \ 0 \ ,  $$
the last equality following from $B=-B^T$. That is, (5.13)
is indeed symmetric under simultaneous identical rotations of the $x$ and $w$ vectors.

\section{Discrete symmetries}

We will now briefly consider, for the sake of completeness, {\it
discrete} symmetries of stochastic differential equations.
Similarly to what happens in the deterministic case (see e.g. \cite{GRo}), the resulting determining equations are in general too difficult to be attacked
except for very simple classes of transformations (e.g. reflections, or
however linear ones). On the other side, they can be used to check
if a given discrete transformation is a symmetry of a given stochastic
differential equation; and/or to determine the stochastic differential
equations which admit a given discrete transformation as a symmetry.

We consider again an Ito equation of the form (2.1); here we will not consider
transformations acting on the $t$ coordinate, and will thus limit to consider the change of coordinates in the $x$ and $w$ spaces given by
$$ y^i = \phi^i (x,t) \ \ , \ \ w^k (t) = R^k_{~p} (t) z^p (t) \ .
\eqno(7.1) $$
As required by lemma 4, we take $R \in O (n)$. We write as usual $A = - (1/2) \s \s^T$.

With the Ito formula we have at once that $y$ obeys the Ito equation
$$ d y^i \ = \ \[ {\pa \phi^i  \over \pa x^j} f^j  \,
- \, A^{jk} {\pa^2 \phi^i  \over \pa x^j \pa x^k} \, + \, {\pa \phi^i
\over \pa t} \] \, d t \ + \ \[ {\pa \phi^i  \over \pa x^j}
\s^j_{~p}  R^p_{~k}  \] \, d z^k \ . \eqno(7.2) $$

\medskip\noindent
{\bf Proposition.} {\it The transformation (6.1) is a symmetry of the Ito equation (2.1) if and only if $\phi , R$ satisfy the {\rm determining equations for discrete symmetries} of an Ito equation:
$$ \begin{array}{l}
{\pa \phi^i (x,t) \over \pa x^j} \, f^j (x,t) \, + \, A^{jk} (x,t)
\, {\pa^2 \phi^i (x,t) \over \pa x^j \pa x^k} \, + \, {\pa \phi^i (x,t)
\over \pa t}\ = \ f^i \( \phi (x,t),t \) \ , \\
{\pa \phi^i (x,t) \over \pa x^j} \, \s^j_{~p} (x,t) \, R^p_{~k} \
= \ \s^i_{~k} \( \phi (x,t) , t \) \ . \end{array} \eqno(7.3) $$}
\medskip

{\bf Example 1.} The simplest case of discrete transformation is
provided by $\phi^i (x,t) = - x^i$, $R= \pm I$. In these cases,
(7.3) reduce to
$$ f^i (x,t) = - f^i (-x,t) \ \ ; \ \ \s (x,t) = \mp \s (-x , t) \ .
\eqno(7.4) $$
\medskip

{\bf Example 2.} For the $n$ independent ``Langevin
oscillators'', see (5.12), it is immediate to check that
(7.4) is satisfied (for $R=-I$).
More generally, (7.3) read now (no sum over $i$,
sum over $j$ when repeated)
$$ a_{(j)} {\pa \phi^i \over \pa x_j} - {s_{(j)}^2 \over 2} {\pa^2
\phi^i \over \pa x^j \pa x^j} \ = \ a_{(i)} \phi^i \ \ ; \ \
s_{(j)} {\pa \phi^i \over \pa x_j} R^j_{~k} \ = \ s_{(i)}
\de^i_{~k} \ . $$
\medskip

{\bf Example 4.} For equations with $\s^i_{~k} (x,t) = s_0 \de^i_{~k} $
(with $s_0$ a real constant), the second of (7.3) always reduces to
$ (\pa \phi^i / \pa x^j) = (R^{-1} )^i_{~j}$. But this implies $\phi$
must be a linear function of the $x$, $\phi^i (x,t) = L^i_{~j} (t) x^j$
with $L$ an orthogonal matrix ($L = R^{-1}$). Thus in this case we are
always reduced to the simple equation
$$ L^i_{~j} \, f^j (x,t) \ = \ f^i (Lx,t) \ \ \ [L \in O(n)] \ . $$

\section{Symmetries of the discrete KPZ equation}

\def\la{\lambda}
\def\La{\Lambda}
\def\th{\theta}
\def\Ga{\Gamma}
\def\lapl{\triangle}
\def\grad{\nabla}
\def\d{{\rm d}}
\def\Gj{{G_{(j)}}}

The KPZ (Kardar-Parisi-Zhang) equation is written, in physicists'
notation, as
$$ {\pa h \over \pa t} \ = \ \nu \lapl h \, + \, {\la \over 2} \, \(
\grad h \)^2 \, + \, \eta \ . \eqno(8.1) $$
Here $h = h(y,t)$ is a scalar variable, representing the height of a
growing (non-overhanging) interface at point $y \in R^n$ at time
$t$, the operators $\grad$ and $\lapl$ represents the gradient and
laplacian with respect to the $y$ variables, $\nu > 0$ and $\la \ge 0$
are real numbers, and $\eta = \eta (y,t)$ is a random field with
$\delta$ spatial correlation function and the statistical properties of
a white noise at each given point $y$. When $\la = 0$ the KPZ reduce to its linear part, which is also known as the Edwards-Wilkinsons equation.

The KPZ is a universal equation describing, under general hypotheses
and up to an asymptotic procedure, the time evolution of a growing interface; see \cite{Bar,HHZ,KPZ} for details.

We will limit ourselves to the one-dimensional case $n=1$; moreover, as our discussion deals only with stochastic ODEs (as opposed to stochastic PDEs), we consider a discretization of the KPZ, obtained by considering a
uniform chain of intersite distance, $(\de y)$; in other words, we
consider a discrete set of points $y_k = k (\de y)$. Thus $h(y,t)$ is
now replaced by a vector variable $x (t)$ with components $x^i (t)$
($i \in {\bf Z}$), where $x^i (t) \approx h (y_i , t)$.

The gradient and Laplacian terms in the KPZ will then be replaced by
$$ (\grad h ) (y_i,t) \approx {x^{i+1} - x^{i-1} \over 2 (\de y) } \ ;
\ (\lapl h) (y_i ,t) \approx {x^{i+1} - 2 x^i + x^{i-1} \over (\de
y)^2} \ . \eqno(8.2) $$
As for $\eta (y,t)$, its discretization will be a standard vector Wiener process
$w (t)$ with component $w^i (t)$ at the point $y_i$.
We also write $\a = \nu [2 (\de y)]^{-1}$ and $\b = \la [2 (\de
y)^2]^{-1}$.

In this way, we replace (8.1) by the Ito equation
$$ \d x^i \ = \ \( \a \[ x^{i+1} - 2 x^i + x^{i-1} \] \ + \ \b \[
x^{i+1} - x^{i-1} \]^2 \) \ \d t \ + \ d w^i (t) \ . \eqno(8.3) $$

It will be convenient to rewrite this in a more abstract notation as
$$ \d x^i \ = \ \[ M^i_{~k} x^k + \Ga^i_{~jk} x^j x^k \] \, \d t \ + \
\d w^i (t) \  \eqno(8.4) $$
(note we have $\s = I$); by comparing (8.3) and (8.4) we have
$$ \begin{array}{l}
M^i_{~k} \ = \  \a \[ \de^i_{~k+1} - 2 \de^i_{~k} + \de^i_{~k-1} \] \\
\Ga^i_{~jk} \ = \ \b \[ \de^i_{~j-1} \de^i_{~k-1}  + \de^i_{~j+1}
\de^i_{~k+1} - \de^i_{~j-1}  \de^i_{~k+1} - \de^i_{~j+1}  \de^i_{~k-1}
\] \ . \end{array} \eqno(8.5) $$
Note that $\Ga^i_{jk} = \Ga^i_{kj}$; it will be convenient to also write $(G_{(j)})^i_k = \Ga^i_{jk}$.

Physically an equation describing a growing interface is required to possess the following symmetries \cite{Bar,HHZ,KPZ}:
(1) time translation; (2) translation in the $h$ direction; (3) space
translations; (4) spatial inversion across any plane and (for $n>1$)
rotations around an axis perpendicular to the $\xb$ plane. Moreover, at the linear level we should also require invariance under: (5) inversion in the $h$ direction.\footnote{Checking that this is the case for the KPZ equation is immediate.}

It is well known that the KPZ is the simplest nonlinear (stochastic PDE) equation possessing these symmetries, and actually it was determined essentially on the basis of this consideration.

We will now check that with our definition of symmetry for a stochastic
differential equation, the discretized KPZ (8.3) does indeed admit
these symmetries. Note that in the discretized case, (1) and (2) remain continuous symmetries, while
(3), (4) and (5) are discrete ones. In particular, (3) is replaced by the
shift of one site along the chain, and (4) reduces to spatial inversion.

\subsection{Continuos symmetries}

We start by analyzing continuos symmetries.
The second of the determining equations (6.2) provides, for $\s = I$,
$$ \xi^i = (1/2) \tau_t x^i - B^i_k x^k + \a^i (t) \ , $$
which we also write as $\xi^i = \La^i_k x^k + \a^i$. Plugging this into the first of (6.2), and separating terms homogeneous of different degree in the $x$, we get (using also symmetry of the $\Ga$ and our shorthand notation)
$$ \begin{array}{l}
\a^i_t - M^i_j \a^j \ = \ 0 \\
{1 \over 2} \[ \tau_{tt} \de^i_j - (B_t)^i_j \] - \tau_t M^i_j + [ \La , M ]^i_j - 2 \a^p \Gj^i_p \ = \ 0 \\
\tau_t \Gj^i_k - \La^i_m \Gj^m_k + \Gj^i_p \La^p_k  \ = \ 0 \end{array} \eqno(8.6) $$

We will not try to determine the most general solution to these, but just check that (1)-(5) are indeed solutions to these equations, i.e. that our definitions correctly identify them as symmetries of the (discretized) KPZ equation.

As for (1), in this case $\tau = 1$, while $\xi^i = 0$, which of course implies $\a$ and $\Gamma$ are also zero, and $B = 0$ as well. It is immediate to check that (8.6) are indeed satisfied in this case.

For the transformation (2) we have $\tau=0$, $B=0$, $\xi^i = 1$; the latter means $\Lambda = 0$, $\a^i = 1$. The first and third of (8.6) are immediately seen to hold, while the second reduces to $\sum_j \Gamma^i_{jk} = 0$; with the explicit expression for $\Gamma$ this is also easily seen to hold, as it reads
$$ \sum_{j=-\infty}^{+\infty} \ \[ \de^{j-1}_{k-1} + \de^{j+1}_{k+1} - \de^{j-1}_{k+1} - \de^{j+1}_{k-1} \] \ . $$

\subsection{Discrete symmetries}

In the case of discrete symmetries (3), (4), (5), we deal with linear transformations, i.e. $x \to y = Fx$, $w \to z = R w$. The relevant equations (7.3) are then
$$ F^i_j f^j \ = \ f^i (Fx) \ \ ; \ \ F^i_j \s^j_p (R^{-1})^p_k \ = \ \s^i_k \ . \eqno(8.7) $$
As $\s = \de$, the second of these actually reads $R = F$. Writing $f$ as in (8.4) and separating linear and quadratic terms in the $x$, the first of (8.7) reads
$$ \begin{array}{l}
\[ F , M \] \ = \ 0 \\
F^i_j \, \Ga^j_{k m} \ = \ \Ga^i_{pq} \, F^p_k \, F^q_m
\end{array} \eqno(8.8) $$
We thus have to check these are satisfied, with $M$ and $\Lambda$ provided by (8.5), for (3),(4),(5).

In the case of translations, (3), we have $F^i_j = \de^i_{j+1}$. Hence
$$ (F M)^i_k = \de^{i-1}_{k+1} - 2 \de^{i-1}_k + \de^{i-1}_{k-1} \ , \ (M F)^i_k = \de^i_{k+2} - 2 \de^i_{k+1} + \de^i_k  \ , $$ and the first of (8.8) is satisfied. Similarly,
$$\begin{array}{l}
F^i_m \Gamma^m_{j k} \ = \ \de^{i-1}_{j-1} + \de^{i-1}_{k-1} + \de^{i-1}_{j+1} \de^{i-1}_{k+1} - \de^{i-1}_{j-1} - \de^{i-1}_{j+1} \de^{i-1}_{k-1} \ , \\
\Gamma^i_{p q} F^p_j F^q_k \ = \ \de^i_j \de^i_k + \de^i_{j+2} \de^i_{k+2} - \de^i_j \de^i_{k+2} - \de^i_{j+2} \de^i_k \ : \end{array} $$
and the second of (8.8) is satisfied as well.

Let us come to (4), i.e. spatial inversion with respect to any site $m$: this is described by $F^i_k = \de^{i-m}_{m-k}$ (no sum on $m$ in this and the following formulas). In this case we have
$$ (FM)^i_j = \de^{i-m}_{m-j-1} - 2 \de^{i-m}_{m-j} + \de^{i-m}_{m-j+1} \ ; \
(MF)^i_j = \de^{i-m-1}_{m-j} - 2 \de^{i-m}_{m-j} + \de^{i-m+1}_{m-j} \ . $$
This shows that the first of (8.8) holds. As for the second, it also holds as it reads
$$\begin{array}{l}
\de^{i-m}_{m-j+1} \de^{i-m}_{m-k+1} + \de^{i-m}_{m-j-1} \de^{i-m}_{m-k-1} - \de^{i-m}_{m-j+1} \de^{i-m}_{m-k-1} - \de^{i-m}_{m-j-1} \de^{i-m}_{m-k+1} \ = \\
= \ \de^{i-m+1}_{m-j} \de^{i-m+1}_{m-k} + \de^{i-m-1}_{m-j} \de^{i-m-1}_{m-k} -
\de^{i-m+1}_{m-j} \de^{i-m-1}_{m-k} - \de^{i-m-1}_{m-j} \de^{i-m+1}_{m-k} \ . \end{array}$$

Finally, let us consider (5): this corresponds to $F = - I$, so obviously $[F,M]=0$. The second of (8.8) reads in this case $\Gamma = 0$, which only holds for $\la = 0$, i.e. in the linear (Edwards-Wilkinson) approximation.

We conclude that our definition correctly captures the symmetry properties of teh KPZ equation.

\section*{Acknowledgement}

This paper was written while the author was a guest at the
DFG-Graduierten\-kol\-leg "Hierarchie und Symmetrie in mathema\-ti\-schen Mo\-del\-len"
at RWTH Aachen. He would like to thank RWTH Aachen, and in particular prof.
S. Walcher, for their hospitality and support.

\end{document}